# Skyrmion Sliding Switch in a 90-nm-Wide Nanostructured Chiral Magnet

*Yaodong Wu[1,2#], Jialiang Jiang[3#], Weiwei Wang[3], Lingyao Kong[3], Shouguo Wang[4], Mingliang Tian[2,3], Haifeng Du[2*], and Jin Tang[3*]*

[1]School of Physics and Materials Engineering, Hefei Normal University, Hefei 230601, China

[2]Anhui Provincial Key Laboratory of Low-Energy Quantum Materials and Devices, High Magnetic Field Laboratory, HFIPS, Chinese Academy of Sciences, Hefei 230031, China

[3]School of Physics and Optoelectronic Engineering, Anhui University, Hefei 230601, China

[4]Anhui Provincial Key Laboratory of Magnetic Functional Materials and Devices, School of Materials Science and Engineering, Anhui University, Hefei 230601, China

\# Y.W. and J.J. contributed equally to this work.

Email: duhf@hmfl.ac.cn; jintang@ahu.edu.cn

# Abstract

Magnetic skyrmions, renowned for their fascinating electromagnetic properties, hold potential for next-generation topological spintronic devices. Recent advancements have unveiled a rich tapestry of 3D topological magnetism. Nevertheless, the practical application of 3D topological magnetism in the development of topological spintronic devices remains a challenge. Here, we showcase the experimental utilization of 3D topological magnetism through the exploitation of skyrmion-edge attractive interactions in 90-nm-wide confined chiral FeGe and CoZnMn magnetic nanostructures. These attractive interactions result in two degenerate equilibrium positions, which can be naturally interpreted as binary bits for a skyrmion sliding switch. Our theory and simulation reveal current-driven spiral motions of skyrmions, governed by the anisotropic gradient of the potential landscape. Our experiments validate the theory that predicts a tunable threshold current density via magnetic field and temperature modulation of the energy barrier. Our results offer an approach for implementing universal on-off switch functions in 3D topological spintronic devices.

Magnetic skyrmions, which are topological particle-like spin vortices, exhibit significant potential as information carriers in the burgeoning realm of topological spintronics.[1-9] Initially conceptualized as two-dimensional spin models[10], recent advancements have significantly broadened the scope of topological spin textures into three-dimensional geometries.[11-15] The discovery of diverse 3D topological spin textures, encompassing skyrmion bundles,[16-18] bobbers,[19] skyrmion-antiskyrmion tubes,[20, 21] skyrmion-bubble tubes,[22, 23] and Hopfions,[24, 25] has profoundly enriched the functionality of topological spintronics.

The intricate 3D spin twists observed in conical magnetizations of chiral magnets lead to distinctive attractive interactions between topological solitons and boundary geometries.[26] These interactions offer promising avenues for the development of innovative nanoelectronic devices. For instance, theoretical simulations indicate that magnetic attractive interactions in 3D confined chiral magnet geometries can naturally induce the formation of binary channels near edges.[27] The binary channels can serve as natural patterning in multi-channel racetrack memories or provide an alternative skyrmion switch. The on/off switching functionality of topological spintronic devices is typically achieved through topological transformations between skyrmions and saturated states with uniformed magnetizations.[28-32] However, due to the distinctly different configurations between skyrmions and saturated states, there are significant energy differences,[30, 31] necessitating asymmetrical physical conditions, such as tailored magnetic fields[28-31] and current profiles[32], for current-controlled skyrmion-ferromagnet transformations. The binary channels facilitate the realization of an alternative on/off switch, i.e. skyrmion sliding on/off switch, relying on current-induced motions of the skyrmion between predefined equilibrium positions. While experimental studies have predominantly focused on the current-driven dynamic motion of skyrmions in open geometries,[16, 33-36] achieving current-driven motions of skyrmions in strongly confined elementary geometries (size ≈ 90 nm) remains challenging.[37, 38]

Here, we delve into the stabilization and current-induced dynamics of a single skyrmion confined within geometrically constrained stripes of chiral magnets. Leveraging 3D topological magnetism, we experimentally discern two degenerate equilibrium positions for skyrmions confined within a nanostructure, stemming from the attractions between the skyrmion and the edges. Furthermore, we achieve the controlled motions of a single skyrmion between these two equilibrium positions utilizing ultrafast single-pulse currents. Our approach, which realizes the on/off switch function through the utilization of 3D topological magnetism, as opposed to the traditional creation/erasion of skyrmions, could inspire advanced designs for topological spintronic devices.

Figure 1a illustrates the schematic design of the skyrmion sliding switch. With a width of 100 nm, it enables the confinement of a single row of skyrmions within the nanostructured FeGe stripe.[39] Due to attractions between the skyrmion and the edge in the conical background magnetization, a twisted, non-axisymmetric skyrmion tube is positioned close to the edge. Our micomagnetic simulations (detail see Supplementary Section I) of the 280-nm long nanostripe reveal two binary equilibrium positions with the lowest total free energy $E$, as shown in Figure 1b. This energy landscape offers a novel switching functionality in topological spintronic devices, facilitating the motion of the skyrmion between the two binary equilibrium positions, $-x_0$ and $x_0$, driven by electrical currents. In this scenario, half of the nanostructured stripe serves as the detection region for on-off devices. Specifically, the skyrmion located at $-x_0$ and $x_0$ represents the binary bits "1" and "0" (Figure 1c), respectively. The on/off function can be effectively realized through the motion of a single skyrmion within confined nanostructures, rather than relying on its creation or deletion. This approach holds significant potential for the development of skyrmion-based random access memories.

We then investigate the dynamics of the single skyrmion driven by an electric current. Initially, the skyrmion is positioned at $-x_0$. Subsequently, an in-plane Zhang-Li spin-transfer torque is applied for a duration of 20 ns.[40] By adopting Thiele's collective coordinate approach and treating the skyrmion as a rigid object,[41, 42] we can deduce the velocity $(v_x, v_y)$ of the motion of the skyrmion, as detailed in Supplementary Section I and presented in Equations (1) and (2).

$$v_x = \frac{\frac{\partial V}{\partial y} - \alpha\eta\frac{\partial V}{\partial x} + u_x(1+\alpha\beta\eta^2)}{1+\alpha^2\eta^2} \quad (1)$$

$$v_y = \frac{-\frac{\partial V}{\partial x} - \alpha\eta\frac{\partial V}{\partial y} - u_x\eta(\alpha-\beta)}{1+\alpha^2\eta^2} \quad (2)$$

Here, $\alpha$ represents the Gilbert damping, $\beta$ is the non-adiabatic parameter, and $\eta$ is the shape factor. The strength of the spin transfer torque is characterized by $\mathbf{u} = \frac{gP\mu_B}{2eM_s}\mathbf{j}$, where $g$, $\mu_B$, $e$, $M_s$, and $P$ stand for the Landé factor, Bohr magneton, electron charge, saturation magnetization, and polarization rate, respectively. In our study, we assume that the current is applied in the $x$-direction, i.e., $\mathbf{u} = (u_x, 0)$. $V$ denotes an effective potential expressed as $V = \frac{\gamma}{4\pi\mu_0 M_s d}E$. $\gamma$ is the gyromagnetic ratio and $d$ is the thickness. In free geometries, the skyrmion undergoes linear motion due to the zero gradients of the potential $V$. However, in confined geometries, the potential $V$ depends on specific locations $(x, y)$ (Figure S1a). Consequently, the skyrmion dynamics in confined geometries exhibit spiral motion dynamics that differ significantly from those in free geometries (Figure S1b and S1c). Using the energy profile as a function of locations $x$ and $y$ based on Equations (1, 2), we can calculate the current-driven dynamic motion trajectories, which align well with the results from micromagnetic simulations[43], as depicted in Figure 1d and 1e. Under the influence of a 20-ns pulsed current with a density of $6.0 \times 10^{10}$ A/m², the skyrmion cannot overcome the energy barrier at $x = 0$ nm and is repelled back to its initial location $-x_0$ (Figure 1d and Supplementary Movie 1). In contrast, a pulsed current with a higher density of $j = 7.0 \times 10^{10}$ A/m² can induce the sliding motion of the skyrmion from the location $-x_0$ to $x_0$ (Figure 1e and Supplementary Movie 2). Therefore, there exists a threshold current density $j_c$ necessary for the motion between two binary equilibrium locations.

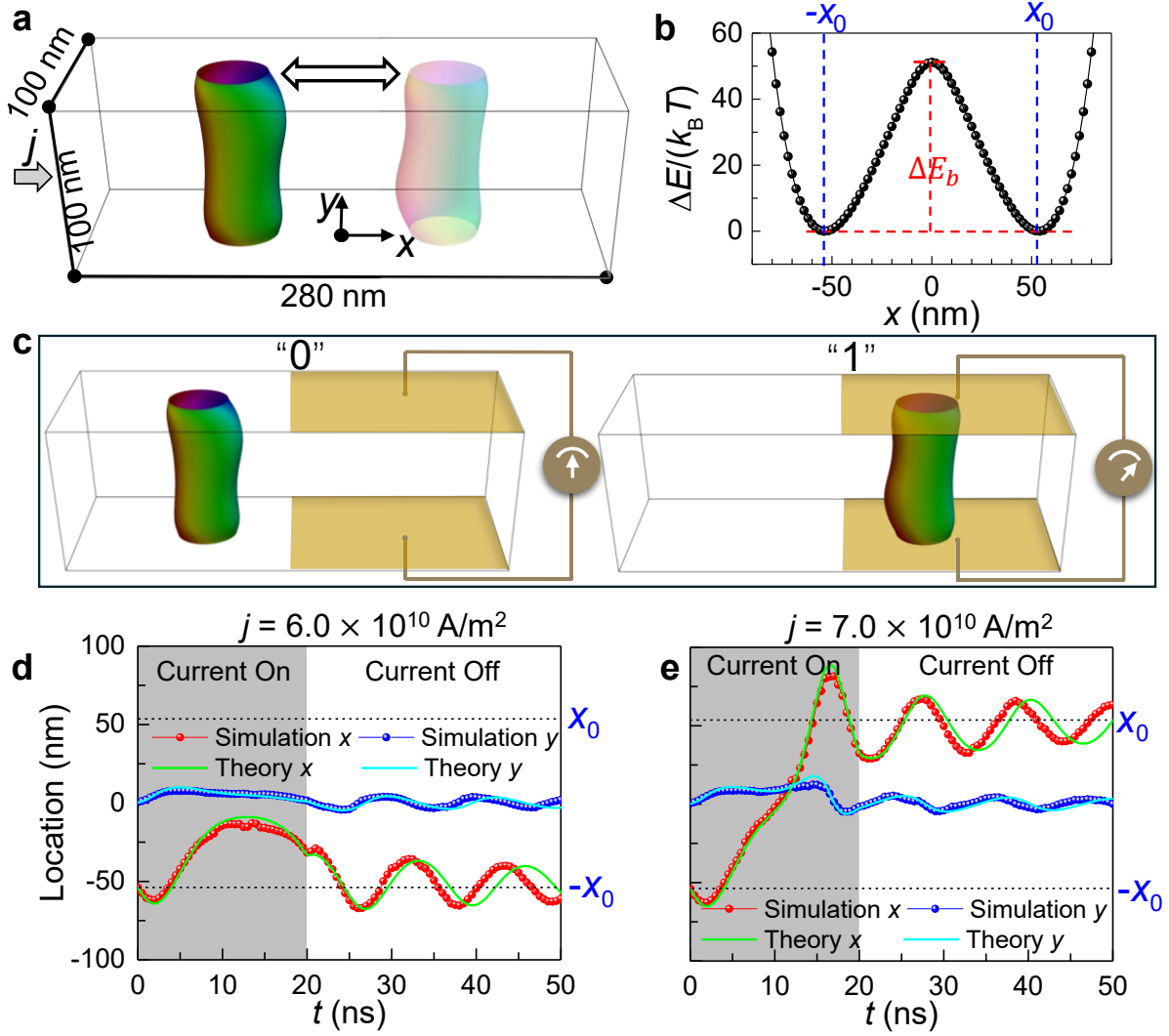

**Figure 1.** Theoretical prediction of the skyrmion sliding switch in a chiral magnetic cuboid. (a) Schematic illustration of the skyrmion sliding device design. (b) Simulated the total free energy difference $\Delta E$ as a function of the skyrmion's position along the $x$-axis. $\Delta E = E - E_{x0}$, $E_{x0}$ is the total free energy for skyrmion located at $-x_0$. Here, we set $T$ = 295 K. $k_B$ is the Boltzmann constant. (c) Schematic design for magnetic random-access memory based on the current-driven sliding motion of a single skyrmion in the nanostructure. The Skyrmion tube located outside and inside of the detection region represents data bits "1" and "0", respectively. (d) Simulated and theoretical time dependence of the skyrmion's position in response to a pulsed current stimulus with a density of $6.0 \times 10^{10}$ A/m². (e) Simulated and theoretical time dependence of the skyrmion's position in response to a pulsed current stimulus with a density of $7.0 \times 10^{10}$ A/m². Magnetic field $B$ is set at 200 mT.

The application of current stimuli can alter the effective potential $V'$. The effective potential $V'$ at the initial location $(-x_0, 0)$ increases as the current density increases (Figure S2). We can estimate the threshold current density $j_c$ required for the skyrmion sliding switch based on the potential balance between the initial location $(-x_0, 0)$ and barrier center $(0, 0)$ (Supplementary Section I):

$$j_c \sim \frac{\gamma e}{2\pi\mu_0 d g P \mu_B \beta \eta} \frac{\Delta E_b}{x_0} \tag{3}$$

Here, $\Delta E_b$ is the energy barrier preventing the skyrmion sliding switch. Taking the exact values for $\Delta E_b$ and $x_0$ at $B = 200$ mT (Figure S3 and S4), the threshold current density $j_c$ estimated according to Equation (3) is about $7.0 \times 10^{10}$ A/m², which agrees well with the value ($6.3 \times 10^{10}$ A/m²) in simulation.

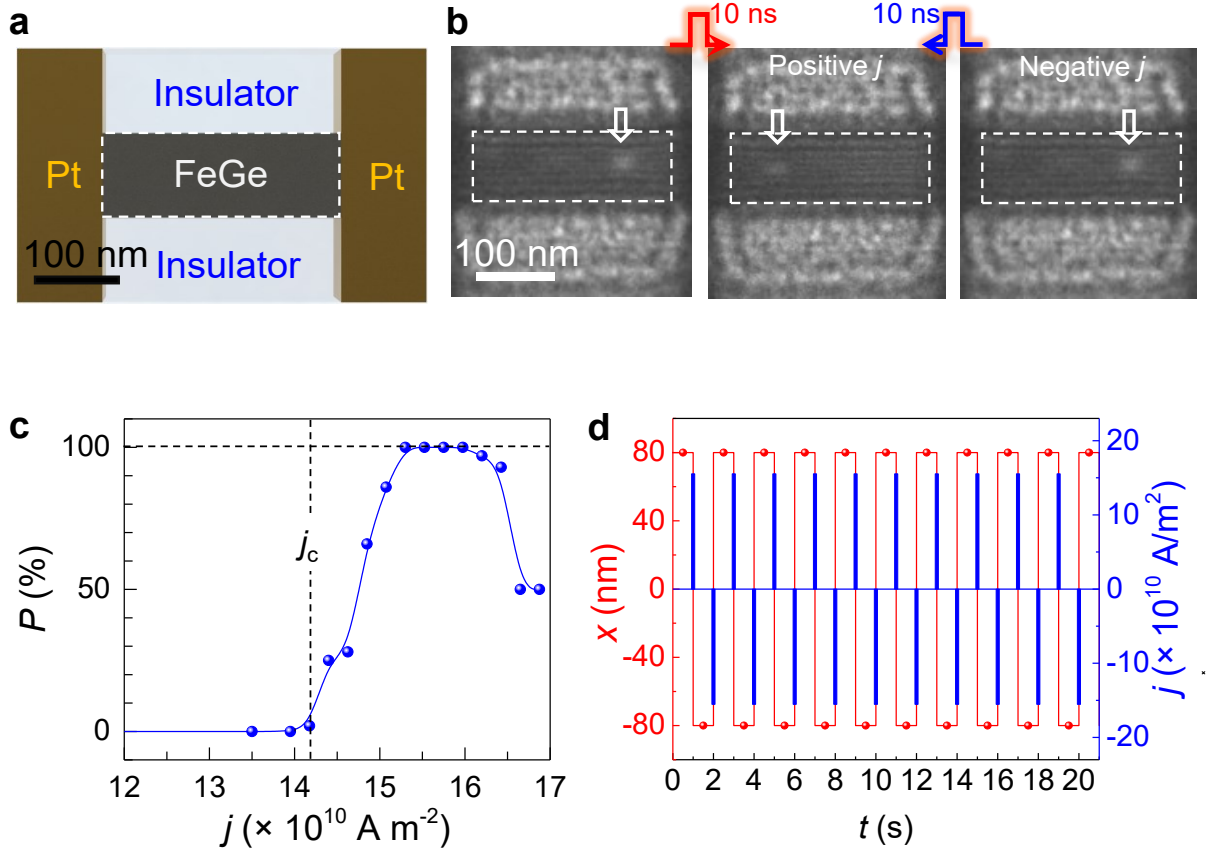

**Figure 2.** Experimental realization of skyrmion sliding switch in a FeGe microdevice. (a) Schematic FeGe microdevice comprising of a nanostructured FeGe stripe encircled by insulator ion-sputtered carbon, connected to two Pt electrodes. (b) Experimental Fresnel images showing a skyrmion motion between $-x_0$ and $x_0$ triggered by a positive and subsequent negative pulsed current. The current density is 15.5×10$^{10}$ A/m². The dashed box indicates the region of the FeGe nanostripe. (c) Dependence of switching probability $P$ on current density $j$. The switching probability is obtained in 100 switching cycles. (d) Reproducible sliding switch of a single skyrmion between two locations, triggered by alternating pulsed currents. The density of single pulsed currents is switched between −15.5 and 15.5 ×10$^{10}$ A/m². Magnetic field $B$ = 175 mT. Pulse duration $w$ is 10 ns.

Figure 2 demonstrates the experimental realization of current-driven skyrmion motion within a confined FeGe nanostripe with a width of ≈ 90 nm and a length of ≈ 280 nm. The FeGe bulk crystals were prepared by the chemical vapor transport method. The magnetic imaging is performed using Fresnel mode of transformation electronic microscopy. Fresnel fringes (the horizontal lines shown in Figure 2b) that appear along the boundary of the geometry due to the edges of the sample are commonly observed under or over-focus conditions. These fringes grow larger as the defocused distance increases.[39, 44] Initially, by increasing the field from 0 to 175 mT at 95 K (Figure S5), we obtain a single skyrmion positioned at $x_0$ = 80 nm, as shown in Figure 2b. A single positive pulsed current along the x-axis with a duration of 10 ns is then applied. When the current density $j$ is below a threshold value $j_c$ (approximately 14.2×10$^{10}$ A/m$^2$), the skyrmion remains static, exhibiting no dynamic responses (Figure 2c and Supplementary Movie 3). However, once the current density surpasses $j_c$, the skyrmion gains the ability to move between locations −$x_0$ and $x_0$ (Supplementary Movies 4 and 5). Notably, the probability of successful switching $P$ increases as the current density increases. Significantly, our findings reveal that for a current density of $j$ = 15.5×10$^{10}$ A/m$^2$, $P$ achieves 100% in at least 100 switching cycles (Figure 2d and Supplementary Movie 5), indicating the high reliability of the skyrmion sliding switch for potential device applications. However, once the current density is above $16.5\times10^{10}$ A/m$^2$, the skyrmion reveals disorder shifts because Joule thermal heating effects become dominant (Fig. S6 and Supplementary Movie 6). The successful switching $P$ decreases to about 50 % for a current density of $j = 16.9\times10^{10}$ A/m$^2$.

It is important to note that the device design for the skyrmion sliding switch (Figure 1a) is not exclusive to FeGe, but it is universally applicable to chiral magnets with bulk Dzyaloshinskii-Moriya interactions (DMIs), including the room-temperature chiral magnet Co$_8$Zn$_{10}$Mn$_2$.[45] Furthermore, we show the feasibility of realizing a current-controlled skyrmion sliding switch operating at room temperature in confined Co$_8$Zn$_{10}$Mn$_2$ nanostripes (Figure S7 and Movie 7). The emergence of binary equilibrium positions is intimately tied to the conical background magnetizations, which arise due to the bulk DMI. We note that such conical background magnetization states may also arise from frustrated interactions in centrosymmetric magnets, such as Gd$_3$Ru$_4$Al$_{12}$.[46] Conversely, for diverse skyrmion varieties, including anisotropic 2D DMI-stabilized antiskyrmions,[47] interface DMI-stabilized Néel-type skyrmions,[48-50] and dipolar-stabilized skyrmions,[51] the background magnetizations consistently exhibit ferromagnetic characteristics. This results in repulsive interactions at the skyrmion and edge.[51, 52] In these skyrmion materials, skyrmions prefer to stay at the center, necessitating the assistance of intricate, additional artificial defects to facilitate the operation of a schematic skyrmion sliding switch.[53]

Figure 3 illustrates the influence of magnetic field strength on the performance of the current-driven skyrmion sliding switch. The binary equilibrium positions within the confined stripe are achieved due to attractive interactions between the skyrmion and the stripe edges within a conical magnetization background. As the magnetic field strengthens, the magnetic background transitions from a conical to a saturated state (Figure 3b), resulting in a change in the skyrmion-edge interactions from attractive to repulsive.[26] Consequently, stable skyrmions located close to the stripe edges are only observable at low magnetic fields. At higher magnetic fields, the equilibrium positions gradually shift towards the center of the nanostripe (Figure 3a and Figure S8). Simulations accurately reproduce this shift of equilibrium positions (Figure 3b and Figure S8). Additionally, as the magnetic field increases, the threshold current density $j_c$ decreases (Figure 3c). This observation can be explained by the modulation of the energy barrier under varying magnetic fields, as depicted in Figure 3d and Figure S3. As the magnetic field rises, the energy barrier decreases, which is consistent with the decrease in the threshold current density predicted by Equation (3) and observed in the simulations (Figure 3d).

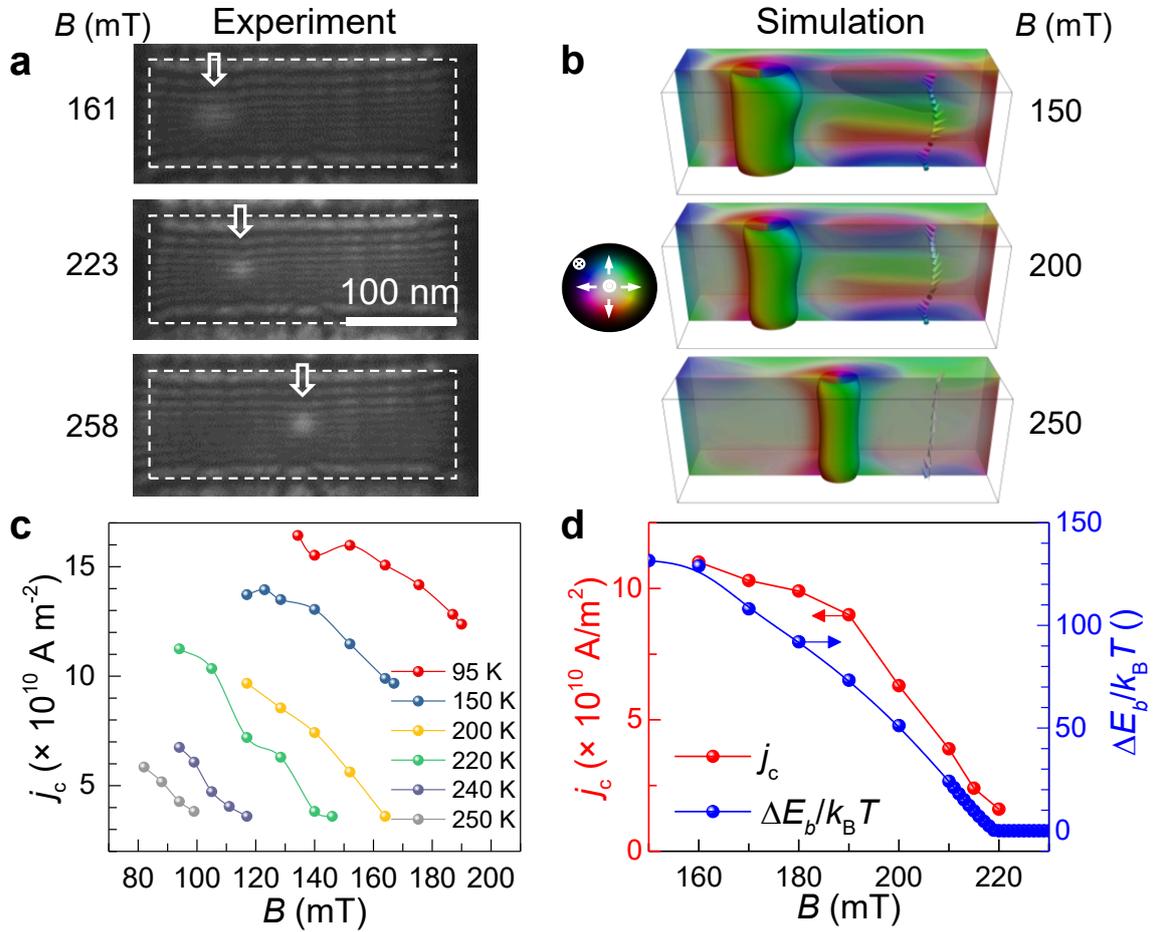

**Figure 3.** Influences of current density $j$, external magnetic field $B$, and temperature $T$ on the sliding switch of single skyrmions. (a) and (b) Skyrmion positioned near to edge to center in the $B$-increasing process revealed in experiments (a) and (b). The color represents magnetization according to the colorwheel in (b). (c) Experimental magnetic field $B$

dependence of critical current density $j_c$ across a temperature range from 95 to 250 K, keeping the pulse duration constant at 10 ns. (d) Simulated dependence of critical current density $j_c$ and energy barrier $\Delta E_b/(k_B T)$ on the magnetic field $B$.

The threshold current density $j_c$ exhibits a decreasing trend as temperature rises (Figure 3c). This trend can be attributed to the reduction in the energy barrier at higher temperatures due to thermal fluctuations (Figure S9). Our simulations further revealed that as the thermal fluctuation fields increase, the skyrmion gradually shifts towards the center at a fixed magnetic field (Figure 4a), resembling the effect of an increased magnetic field at a fixed temperature. Furthermore, our simulations align with experimental findings, demonstrating a decrease in the threshold current density as temperature increases (Figure 4b). We have also investigated the dependence of the threshold current density on pulse duration ($w$). Our experiments indicate that the threshold current density $j_c$ decreases from $16.6 \times 10^{10}$ A/m² for a pulse duration of 5 ns to $4.9 \times 10^{10}$ A/m² for a pulse duration of 200 ns (Figure 4c). This decrease can be attributed to the significant Joule heating effects induced by longer pulse durations. The transient temperature increase contributes to the reduction of the energy barrier, ultimately leading to a decrease in the threshold current density.

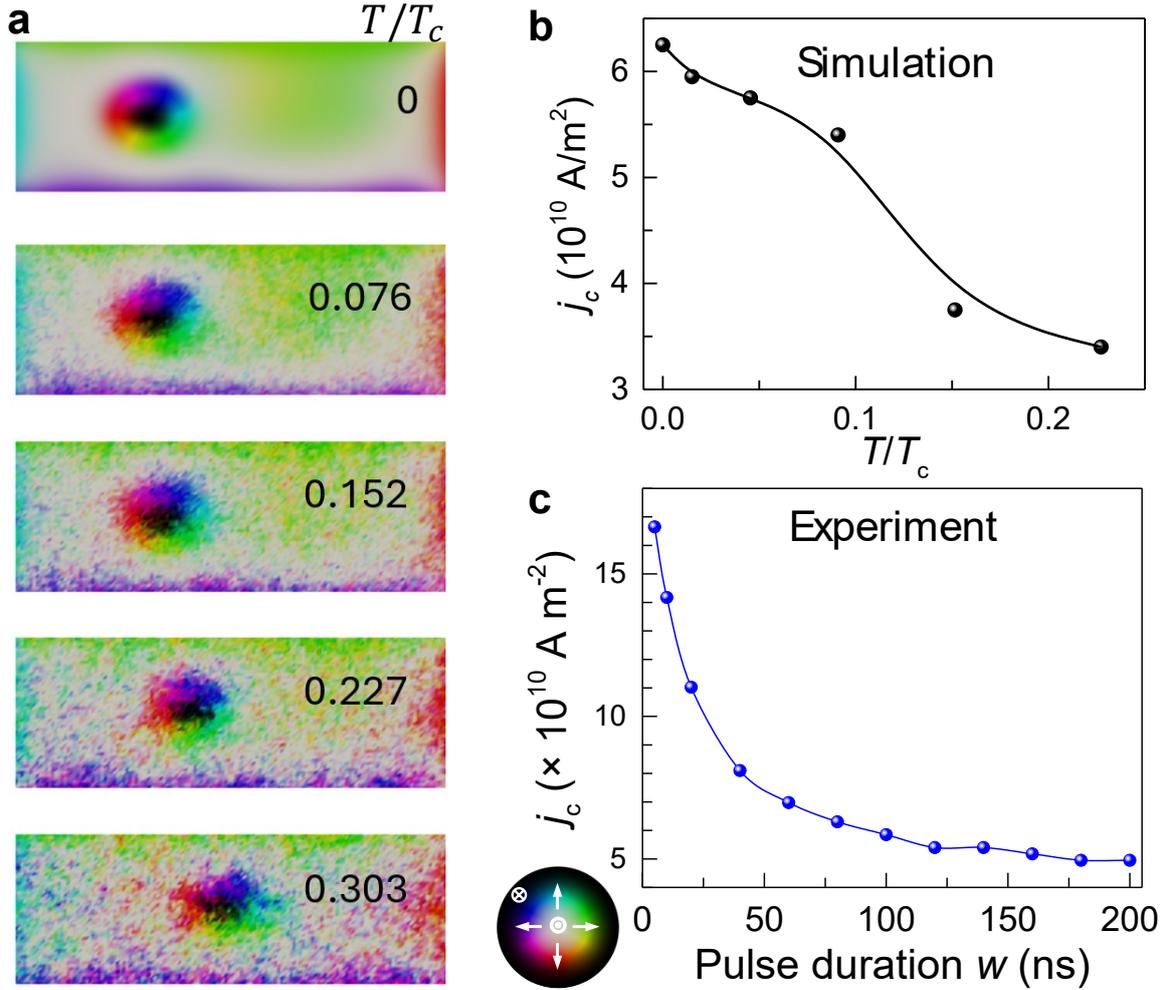

**Figure 4.** Effect of pulse duration *w* on skyrmion sliding switch. (a) Simulated threshold current density $j_c$ as a function of temperature. Here, $T_c$ is the Curie temperature at $B = 200$ mT. The color represents magnetization according to the colorwheel. (b) Simulated temperature dependence of skyrmion shifts at a fixed field 200 mT. (c) Experimental pulse duration *w* dependence of threshold current density $j_c$ at 200 mT and 95 K.

We next compare the minimum energy consumption required for skyrmion sliding switches and skyrmion creation/deletion, both of which are relevant for skyrmion-based on/off operations. The minimum energy consumption ($E_0$) is calculated using the equation $E_0 = I^2Rw$, where $I$ is the critical current required to trigger the on/off operation, and $R$ is the resistance of the magnetic cell. For skyrmion sliding switches in FeGe cuboids, the minimum energy consumption falls within the range of 0.1–1 pJ/bit (Fig. S10). We also evaluate the energy cost of writing and deleting individual skyrmions in confined $Fe_3Sn_2$ nanostructures.[14] The minimum energy consumption for writing and deleting a single skyrmion is approximately 4.5 pJ/bit ($I \approx 5.8$ mA, $R \approx 13.3$ Ω, and $w = 10$ ns) and 31.5 pJ/bit ($I \approx 15.4$ mA, $R \approx 13.3$ Ω, and $w = 10$ ns), respectively[14]—values significantly higher than those of the sliding switch mechanism. The size of a skyrmion is inversely proportional to the strength of DMI.[54] This relationship implies that the core components of skyrmion-based sliding switches—currently scaled to sub-90 nm dimensions—could be miniaturized further in magnetic systems with enhanced DMI. Moreover, reducing the size of these nanostructured cells may lower the minimum energy required for their operation, enabling more energy-efficient spintronic devices.

It is important to note that the thickness of the nanostructured cuboid can significantly alter its energy landscape. Simulations using the magnetic parameters of FeGe reveal that binary equilibrium states vanish when the thickness falls below approximately 30 nm (Fig. S11). For sufficiently thin films, in-plane shape anisotropy—driven by dipolar-dipolar interactions—becomes dominant, leading to the disappearance of conical background magnetization states. The demagnetization energy confines the skyrmion to the center of the 20-nm-thick cuboid across a broad range of applied fields.

In summary, we have validated both theoretically and experimentally the concept of a current-driven skyrmion sliding switch in confined chiral magnetic nanostructures. The attractive skyrmion-edge interactions, stemming from 3D conical spin modulations, naturally create two equilibrium positions in the simple rectangular nanostructure, enabling binary information coding. The threshold current density is established based on the ratio between the energy barrier and the equilibrium locations, which can be effectively adjusted by varying external magnetic fields and temperatures. Our findings demonstrate that 3D topological magnetism can be harnessed to develop innovative functionalities in spintronic devices, including random access memory based on the skyrmion sliding switch and binary channels for skyrmion racetrack memory. The high repeatability and relatively low energy consumption of the current-driven skyrmion sliding switch hold promise for advancing topological spintronic devices with high performance.

## METHODS

**Preparation of Chiral Magnets.** $B_{20}$ type FeGe single crystals were grown by chemical vapor transport method with a mixture of stoichiometric iron (Alfa Aesar, purity > 99.9%), germanium (Alfa Aesar, purity > 99.99%), and transport agent $I_2$. The B20 type FeGe crystallizes in a temperature gradient from 560 °C to 500 °C, with a nearly pyramidal shape for space group $P2_13$. Polycrystalline samples of $Co_8Zn_{10}Mn_2$ crystals were synthesized by a high-temperature reaction method. Stoichiometric cobalt (> 99.9%), zinc (> 99.99%), and manganese (> 99.95%) were mixed into a quartz tube and sealed under vacuum, heated to 1273 K for 24 h. Then slowly cooled to 1198 K, and maintained at this temperature for more than 72 h. After that, put the tube in cold water to quench. Finally, a spherical $Co_8Zn_{10}Mn_2$ alloy with a metallic luster was obtained.

**Fabrication of Micro-devices.** Thin FeGe and $Co_8Zn_{10}Mn_2$ thin plates of thickness of $t \approx$ 150 nm for TEM magnetic imaging were fabricated by the lift-out method using a focused-ion beam and scanning electron microscopy dual beam system (Helios Nanolab 600i, FEI) in combination with a gas injection system and a micromanipulator (Omniprobe 200+, Oxford).[55]

**TEM Measurements.** Magnetic imaging was carried out using a TEM instrument (Talos F200X, FEI) operated at 200 kV in the Lorentz Fresnel mode. The objective lens is switched off to provide a field-free condition. A single-tilt liquid-nitrogen specimen holder (Model 616.6 cryotransfer holder, Gatan) was used with a temperature range from 95 to 300 K. The current pulses were provided by a voltage source (AVR-E3-B-PN-AC22, Avtech

Electrosystems Ltd.) and were set to be a frequency of 1 Hz.

**Micromagnetic Simulations.** Micromagnetic simulations were performed using the MuMax3.[43] The total free energy terms are written as:

$$\varepsilon = \int_{V_s} \{\varepsilon_{ex} + \varepsilon_{DMI} + \varepsilon_{zeeman} + \varepsilon_{dem}\} d\boldsymbol{r} \quad (4)$$

Here, exchange energy $\varepsilon_{ex} = A(\partial_x \mathbf{m}^2 + \partial_y \mathbf{m}^2 + \partial_z \mathbf{m}^2)$, DMI energy $\varepsilon_{DMI} = D_{dmi} \mathbf{m} \cdot [\nabla \times \mathbf{m}]$, Zeeman energy $\varepsilon_{zeeman} = -M_s \mathbf{B_{ext}} \mathbf{m}$, and demagnetization energy $\varepsilon_{dem} = -\frac{1}{2} M_s \mathbf{B_d} \mathbf{m}$. Here $\mathbf{m} \equiv \mathbf{m}(x,y,z)$ is the normalized units continuous vector field that represents the magnetization $\mathbf{M} \equiv M_s \mathbf{m}(x,y,z)$. $A$, $D_{dmi}$, and $M_s$ are the exchange interaction, DMI interaction, and saturation magnetization, respectively. $\mathbf{B}_d$ is the demagnetizing field. We set a typical value for saturation magnetization $M_s$ = 384 kA m$^{-1}$ for FeGe[26]. The exchange interaction $A_{ex}$ = 3.25 pJ/m is determined from the fit to field-dependence of magnetization evolution[26]. DMI interaction $D_{dmi} = 4\pi A/L_D$ = 5.834 mJ m$^{-2}$ is obtained from zero-field spin spiral period $L_D$ = 70 nm[26]. We set the cell size as 4 × 4 × 3 nm$^3$. We obtained the equilibrium spin configurations using the conjugate-gradient method.

A Zhang-Li spin transfer torque (STT) is considered for simulating current-driven dynamics:

$$\varepsilon_{ZL} = \frac{1}{1+\alpha^2}\{(1+\beta\alpha)\mathbf{m} \times [\mathbf{m} \times (\mathbf{u} \cdot \nabla)\mathbf{m}] + (\beta - \alpha)\mathbf{m} \times (\mathbf{u} \cdot \nabla)\mathbf{m}\} \quad (5)$$

The strength of spin transfer torque is characterized by $\mathbf{u} = \frac{gP\mu_B}{2eM_s}\mathbf{j}$ with parameters $\mathbf{j}$, $g$, $\mu_B$, $e$, $M_s$, and $P$ are current, the Landé factor, Bohr magneton, electron charge, saturation magnetization, and polarization rate, respectively. Here, we set $g = 2$, $\beta = 0.1336$, $\alpha = 0.0167$ according to a previous study[56].

The finite temperature was provided by the thermal fluctuation filed $\vec{B}_{therm} = \vec{\eta}(\text{step})\sqrt{\frac{2\mu_0 \alpha k_B T}{B_{sat}\gamma_{LL}\Delta V \Delta t}}$, where $k_B$, $T$, $B_{sat}$, $\gamma_{LL}$, $\Delta V$, and $\Delta t$ are Boltzmann constant, temperature, saturation magnetization, gyromagnetic ratio, cell volume, and time step, respectively. $\vec{\eta}$(step) is a random vector changed after every time step.[57]

**ASSOCIATED CONTENT**

**Supporting Information**

The Supporting Information is available free of charge at XX.

**Figure S1:** Energy diagram for the skyrmion located in the nanostructure. **Figure S2:** Diagram of the effective potential $V'$ during the application of current. **Figure S3:** Location $x$ dependence of energy differences the total free energy difference $\Delta E$ as a function of the skyrmion's position along the $x$-axis. **Figure S4:** Magnetic field $B$ dependence of threshold current density $j_c$ required for the skyrmion sliding switch according to Equation (S10). **Figure S5:** Magnetic field B dependence magnetic evolution at 95 K. Disorder skyrmion dynamic shift triggered by alternating pulsed currents. **Figure S6:** Experimental realization of skyrmion sliding switch in $Co_8Zn_{10}Mn_2$ chiral nanostructure at room temperature. **Figure S7:** **Figure S8:** Experimental and simulated magnetic field $B$ dependence of skyrmion equilibrium locations ($x_0$) in the FeGe chiral nanostructure. **Figure S9:** Simulated temperature dependence of out-of-plane magnetization $m_z$ at $B = 200$ mT applied along the $z$-axis. **Figure S10:** Experimental minima energy consumption. **Figure S11:** Simulated effect of thickness on the energy landscape of nanostructured cuboid. **Movie 1:** Simulated current-driven dynamics of a single skyrmion in the FeGe nanostripe. **Movie 2:** Simulated current-driven dynamics of a single skyrmion in the FeGe nanostripe. **Movie 3:** Experimental dynamics of a single skyrmion in the FeGe nanostripe driven by single pulsed currents. **Movie 4:** Experimental dynamics of a single skyrmion in the FeGe nanostripe driven by single pulsed currents. **Movie 5:** Experimental dynamics of a single skyrmion in the FeGe nanostripe driven by single pulsed currents. **Movie 6:** Experimental disorder dynamics of a single skyrmion in the FeGe nanostripe driven by single pulsed currents with a high density. **Movie 7:** Experimental

dynamics of a single skyrmion in the $Co_8Zn_{10}Mn_2$ nanostripe driven by single pulsed currents at room temperature, 295 K.

## AUTHOR INFORMATION

**Corresponding Author**

Haifeng Du, email: duhf@hmfl.ac.cn;

Jin Tang, email: jintang@ahu.edu.cn.

**Author Contributions**

Y.W. and J.J. contributed equally to this work. H.D. and J.T. supervised the project. J.T. conceived the idea and designed the experiments. J.T. fabricated the microdevices and performed TEM measurements with the help of Y.W. and J.J.. Y.W., L.K., and J.T. performed the simulations. J.T., Y.W., and H.D. wrote the manuscript with input from all the authors. S.W. and M.T. discussed the results. All authors contributed to the manuscript.

**Notes**

The authors declare no competing financial interest

## ACKNOWLEDGMENTS

This work was supported by the National Key R&D Program of China, Grant No. 2024YFA1611303; the Natural Science Foundation of China, Grants No. 52325105, 12422403, U24A6001, 12174396, 12104123, and 12241406; the Anhui Provincial Natural Science Foundation, Grant No. 2308085Y32; the Natural Science Project of Colleges and Universities in Anhi Province, Grants No. 2022AH030011 and 2024AH030046; Anhui Province Excellent Young Teacher Training Project Grant No. YQZD2023067; the 2024 Project of GDRCYY (No. 217, Yaodong Wu); the China Postdoctoral Science Foundation Grant No. 2023M743543; CAS Project for Young Scientists in Basic Research, Grant No.

YSBR-084; and Systematic Fundamental Research Program Leveraging Major Scientific and Technological Infrastructure, Chinese Academy of Sciences, Grant No. JZHKYPT-2021-08.

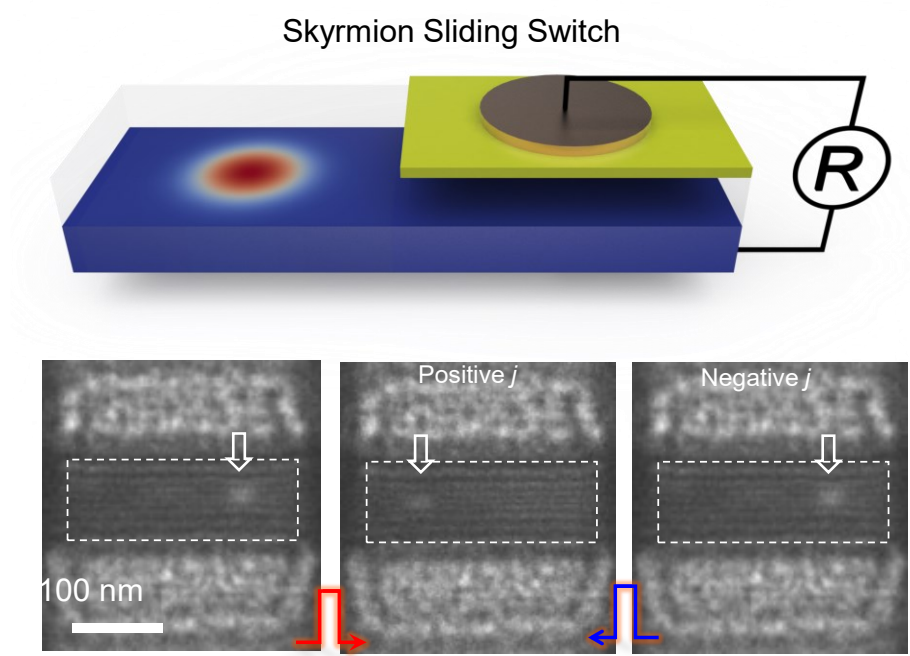

**For Table of Contents Only**